\documentclass{article}
\usepackage{amssymb}
\usepackage{amsfonts}
\usepackage{amsmath}
\usepackage{graphicx}

\begin{document}

\title{Exact nonlinear fourth-order equation for two coupled nonlinear
oscillators: metamorphoses of resonance curves}
\author{Jan Kyziol$^{1)}$, Andrzej Okninski$^{2)}$ \\
Department of Mechatronics and Mechanical Engineering$^{1)}$, \\
Department of Management and Computer Modelling$^{2)}$, \\
Politechnika Swietokrzyska, Al. 1000-lecia PP7, \\
25-314 Kielce, Poland}
\maketitle

\begin{abstract}
We study dynamics of two coupled periodically driven oscillators. The
internal motion is separated off exactly to yield a nonlinear fourth-order
equation describing inner dynamics. Periodic steady-state solutions of the
fourth-order equation are determined within the
Krylov-Bogoliubov-Mitropolsky approach -- we compute the amplitude profiles,
which from mathematical point of view are algebraic curves.

In the present paper we investigate metamorphoses of amplitude profiles
induced by changes of control parameters near singular points of these
curves. It follows that dynamics changes qualitatively in the neighbourhood
of a singular point.
\end{abstract}

\section{Introduction}

In this work we study dynamics of two coupled oscillators, one of which is
driven by an external periodic force. Equations governing dynamics of such
system are of form:%
\begin{equation}
\left. 
\begin{array}{l}
m_{1}\ddot{x}_{1}-V_{1}\left( \dot{x}_{1}\right) -R_{1}\left( x_{1}\right)
+V_{2}\left( \dot{x}_{2}-\dot{x}_{1}\right) +R_{2}\left( x_{2}-x_{1}\right)
=f\cos \left( \omega t\right) \\ 
m_{2}\ddot{x}_{2}-V_{2}\left( \dot{x}_{2}-\dot{x}_{1}\right) -R_{2}\left(
x_{2}-x_{1}\right) =0%
\end{array}%
\right\}  \label{model1}
\end{equation}%
where $R_{1}$, $V_{1}$ and $R_{2}$, $V_{2}$ are nonlinear elastic restoring
force and nonlinear force of internal friction for mass $m_{1}$ and mass $%
m_{2}$, respectively. Dynamic vibration absorber, consisting of a (generally
small) mass $m_{2}$, attached to the primary vibrating system of (typically
larger) mass $m_{1}$ is a generic mechanical model described by (\ref{model1}%
) \cite{DenHartog1985,Oueini1999}.

We shall consider a special case:%
\begin{equation}
R_{1}\left( x_{1}\right) =-\alpha _{1}x_{1},\ V_{1}\left( \dot{x}_{1}\right)
=-\nu _{1}\dot{x}_{1}.  \label{simplified}
\end{equation}

Dynamics of coupled periodically driven oscillators is very complicated \cite%
{Szemplinska1990,Awrejcewicz1991,Kozlowski1995,Janicki1995,Kuznetsov2009,Awrejcewicz2012}%
. Starting from equations (\ref{model1}), (\ref{simplified}) we derived the
exact fourth-order nonlinear equation for internal motion as well as
approximate second-order effective equation \cite{Okninski2005,Okninski2006}
(this approximation performs well for $\tfrac{m_{2}}{m_{1}}\ll 1$). Applying
the Krylov-Bogoliubov-Mitropolsky (KBM) method to the effective equation we
have computed and studied the corresponding nonlinear resonances. More
exactly, we investigated the amplitude profiles (resonance curves) $A\left(
\omega \right) $, i.e. dependence of the amplitude on the frequency $\omega $%
, given implicitly by the KBM method. Metamorphoses of the resonance curves $%
A\left( \omega \right) $ induced by changes of the control parameters,
leading to new nonlinear phenomena, have been studied within the theory of
algebraic curves -- they occur in the neighbourhoods of singular points of $%
A\left( \omega \right) $ \cite{Kyziol2011,Kyziol2012a,Kyziol2012b}.

In the present paper we study the exact fourth-order equation for internal
motion. It turns out that the KBM method can be applied to yield equation
defining the amplitude profile $A\left( \omega ;a,b,\ldots \right) $
implicitly, where $a,b,\ldots $ are some parameters. This resonance curve is
more complicated then in the case of effective equation and hence more
complicated metamorphoses are possible. The aim of the present paper is to
explore these possibilities.

The paper is organized as follows. In the next Section the exact 4th-order
equation for the internal motion in non-dimensional form is presented. In
Section 3 equation for the resonance curves $A\left( \omega \right) $ is
derived from the exact fourth-order equation for internal motion via the
Krylov-Bogoliubov-Mitropolsky approach. In Section 4 the theory of algebraic
curves is used to compute singular points on the exact equation amplitude
profiles - metamorphoses of amplitude profiles occur in neighbourhoods of
such points. In Section 5 examples of analytical and numerical computations
are presented for the 4th-order equation equation. Our results are
summarized in the last Section.

\section{Exact equation for internal motion}

In new variables, $x\equiv x_{1}$, $y\equiv x_{2}-x_{1}$, equations (\ref%
{model1}), (\ref{simplified}) can be written as:

\begin{equation}
\left. 
\begin{array}{l}
m\ddot{x}+\nu \dot{x}+\alpha x+V_{e}\left( \dot{y}\right) +R_{e}\left(
y\right) =f\cos \left( \omega t\right) \\ 
m_{e}\left( \ddot{x}+\ddot{y}\right) -V_{e}\left( \dot{y}\right)
-R_{e}\left( y\right) =0%
\end{array}%
\right\} ,  \label{model2}
\end{equation}%
where $m\equiv m_{1}$, $m_{e}\equiv m_{2}$, $\nu \equiv \nu _{1}$, $\alpha
\equiv \alpha _{1}$, $V_{e}\equiv V_{2}$, $R_{e}\equiv R_{2}$.

Adding equations (\ref{model2}) we obtain important relation between
variables $x$ and $y$:

\begin{equation}
M\ddot{x}+\nu _{1}\dot{x}+\alpha _{1}x+m_{e}\ddot{y}=f\cos \left( \omega
t\right) ,  \label{relation}
\end{equation}%
where $M=m+m_{e}$.

We can eliminate variable $x$ in (\ref{model2}) to obtain the following
exact equation for relative motion: 
\begin{equation}
\left( M\tfrac{d^{2}}{dt^{2}}+\nu \tfrac{d}{dt}+\alpha \right) \left( \mu 
\ddot{y}-V_{e}\left( \dot{y}\right) -R_{e}\left( y\right) \right) +\epsilon
m_{e}\left( \nu \tfrac{d}{dt}+\alpha \right) \ddot{y}=F\cos \left( \omega
t\right) ,  \label{4th-a}
\end{equation}%
where $F=m_{e}\omega ^{2}f$, $\mu =mm_{e}/M$ and $\epsilon =m_{e}/M$ is a
nondimensional parameter \cite{Okninski2005,Okninski2006}, see also Ref. 
\cite{Starosta2011} where separation of variables for a more general system
of coupled equations was described. Equations (\ref{4th-a}), (\ref{relation}%
) are equivalent to the initial equations (\ref{model1}), (\ref{simplified}).

For small $\epsilon $ we can reject the term proportional to $\epsilon $ to
obtain the approximate (effective) equation which can be integrated partly
to yield the effective equation:%
\begin{equation}
\mu \ddot{y}-V_{e}\left( \dot{y}\right) -R_{e}\left( y\right) =\tfrac{%
-m_{e}\omega ^{2}f}{\sqrt{M^{2}\left( \omega ^{2}-\frac{\alpha }{M}\right)
^{2}+\nu ^{2}\omega ^{2}}}\cos \left( \omega t+\delta \right) .  \label{eff}
\end{equation}%
In what follows we shall assume%
\begin{equation}
R_{e}\left( y\right) =-\alpha _{e}y-\gamma _{e}y^{3},\quad V_{e}\left( \dot{y%
}\right) =-\nu _{e}\dot{y}+\lambda _{e}\dot{y}^{3}.  \label{Re-Ve}
\end{equation}%
This model was also investigated in \cite{Awrejcewicz2012} where limiting
phase trajectories approach was used.

In this work we shall investigate the exact equation (\ref{4th-a}). We write
Eqns. (\ref{4th-a}), (\ref{Re-Ve}) in nondimensional form. Introducing
nondimensional time $\tau $ and rescaling variable $y$:

\begin{equation}
\tau =t\bar{\omega},\ z=y\sqrt{\frac{\gamma _{e}}{\alpha _{e}}},
\label{Ndim1}
\end{equation}%
where:%
\begin{equation}
\bar{\omega}=\sqrt{\frac{\alpha _{e}}{\mu }},  \label{omega}
\end{equation}%
we get:%
\begin{equation}
\mathcal{\hat{L}}\left( \tfrac{d^{2}z}{d\tau ^{2}}+h\tfrac{dz}{d\tau }%
-b\left( \tfrac{dz}{d\tau }\right) ^{3}+z+z^{3}\right) +\kappa \left( H%
\tfrac{d}{d\tau }+a\right) \tfrac{d^{2}z}{d\tau ^{2}}=\tfrac{\kappa }{\kappa
+1}G\Omega ^{2}\cos \left( \Omega \tau \right)  \label{4th-b}
\end{equation}%
where $\mathcal{\hat{L}}$ is a linear operator:%
\begin{equation}
\mathcal{\hat{L}}=\tfrac{d^{2}}{d\tau ^{2}}+H\tfrac{d}{d\tau }+a,  \label{L}
\end{equation}%
and nondimensional constants are given by:%
\begin{equation}
h=\frac{\nu _{e}}{\mu \bar{\omega}},\ b=\frac{\lambda _{e}}{\gamma _{e}}\bar{%
\omega}^{3},\ H=\dfrac{\nu }{M\bar{\omega}},\ \Omega =\frac{\omega }{\bar{%
\omega}},\ G=\frac{1}{\alpha _{e}}\sqrt{\frac{\gamma _{e}}{\alpha _{e}}}f,\
\kappa =\frac{m_{e}}{m},\ a=\dfrac{\alpha \mu }{\alpha _{e}M}.  \label{Ndim2}
\end{equation}

\section{Nonlinear resonances via Krylov-Bogoliubov-Mitropolsky method}

We apply the Krylov-Bogoliubov-Mitropolsky (KBM) perturbation approach \cite%
{Nayfeh1981} to the exact nonlinear fourth-order equation (\ref{4th-b})
describing internal motion of the small mass. The equation (\ref{4th-b}) is
written in the following form:%
\begin{equation}
\mathcal{\hat{L}}\left( \tfrac{d^{2}z}{d\tau ^{2}}+\Omega ^{2}z\right)
+\varepsilon \left( \sigma \mathcal{\hat{L}}z+g\left( z,\dot{z}\right)
\right) =0,  \label{prepared}
\end{equation}%
where $\mathcal{\hat{L}}$ is defined in (\ref{L}) and $\varepsilon g\left( z,%
\dot{z}\right) $ is given by:%
\begin{equation}
\varepsilon g=\mathcal{\hat{L}}\left( h\tfrac{dz}{d\tau }-b\left( \tfrac{dz}{%
d\tau }\right) ^{3}+z+z^{3}\right) -\Theta ^{2}\mathcal{\hat{L}}z+\kappa
\left( H\tfrac{d}{d\tau }+a\right) \tfrac{d^{2}z}{d\tau ^{2}}-\tfrac{\kappa 
}{\kappa +1}G\Omega ^{2}\cos \left( \Omega \tau \right) ,  \label{g}
\end{equation}%
and%
\begin{equation}
\Theta ^{2}-\Omega ^{2}=\varepsilon \sigma .  \label{Omega}
\end{equation}%
Equation (\ref{prepared}) was prepared in such way that for $\varepsilon =0$%
\ the general solution, $z\left( \tau \right) =A\cos \left( \Omega \tau
+\varphi \right) $\ $+C\exp \left( -\frac{1}{2}\left( H-\sqrt{\Delta }%
\right) \tau \right) +D\exp \left( -\frac{1}{2}\left( H+\sqrt{\Delta }%
\right) \tau \right) $, $\Delta =H^{2}-4a$, with constant and arbitrary $A$, 
$\varphi $, $C$, $D$\ and, moreover, the solution for $H,a>0$\ does not
contain secular terms and $z\left( \tau \right) \longrightarrow A\cos \left(
\Omega \tau +\varphi \right) $ for $\tau \longrightarrow \infty .$

We shall now look for $1:1$ resonance using the KBM method. For small
nonzero $\varepsilon $ the solution of Eqns.(\ref{prepared})~--~ (\ref{Omega}%
) and (\ref{Re-Ve}) is sought in form:%
\begin{equation}
z=A\cos \left( \Omega \tau +\varphi \right) +\varepsilon z_{1}\left(
A,\varphi ,\tau \right) +\ldots  \label{sol1:1}
\end{equation}%
with slowly varying amplitude and phase:%
\begin{eqnarray}
\dfrac{dA}{d\tau } &=&\varepsilon M_{1}\left( A,\varphi \right) +\ldots ,
\label{A} \\
\dfrac{d\varphi }{d\tau } &=&\varepsilon N_{1}\left( A,\varphi \right)
+\ldots .  \label{phi}
\end{eqnarray}

Computing now derivatives of $z$ from Eqns.(\ref{sol1:1}), (\ref{A}), (\ref%
{phi}) and substituting to Eqns.(\ref{prepared})~--~ (\ref{Omega}), (\ref%
{Re-Ve}) and eliminating secular terms and demanding $M_{1}=0$, $N_{1}=0$ we
obtain the following equations for the amplitude and phase of steady states: 
\begin{subequations}
\label{STEADY}
\begin{align}
-\Omega A\left( p\Omega ^{2}-q\right) +\tfrac{3}{4}\Omega A^{3}\left(
H-ba\Omega ^{2}+b\Omega ^{4}\right) +\tfrac{\kappa }{\kappa +1}G\Omega
^{2}\sin \varphi & =0,  \label{st1} \\
-A\left( \Omega ^{4}-r\Omega ^{2}+a\right) -\tfrac{3}{4}A^{3}\left( a-\Omega
^{2}+Hb\Omega ^{4}\right) +\tfrac{\kappa }{\kappa +1}G\Omega ^{2}\cos
\varphi & =0.  \label{st2}
\end{align}%
where $p=h+H\left( \kappa +1\right) $, $q=ah+H$, $r=hH+a\left( \kappa
+1\right) +1$.

Solving the system of equations (\ref{STEADY}) we get the implicit
expressions for the amplitude $A\left( \Omega \right) $ and the phase $%
\varphi \left( \Omega \right) $: 
\end{subequations}
\begin{subequations}
\label{AF}
\begin{eqnarray}
A\left( \Omega \right)  &=&\dfrac{\kappa G}{\kappa +1}\dfrac{\Omega ^{2}}{%
\sqrt{C^{2}+D^{2}}},  \label{AOmega} \\
\tan \varphi \left( \Omega \right)  &=&\dfrac{C}{D},  \label{phase} \\
C &=&\Omega A\left( \Omega \right) \left( p\Omega ^{2}-q\right) -\tfrac{3}{4}%
\Omega A^{3}\left( \Omega \right) \left( H-ba\Omega ^{2}+b\Omega ^{4}\right) 
\label{C} \\
D &=&A\left( \Omega \right) \left( \Omega ^{4}-r\Omega ^{2}+a\right) +\tfrac{%
3}{4}A^{3}\left( \Omega \right) \left( a-\Omega ^{2}+Hb\Omega ^{4}\right) 
\label{D}
\end{eqnarray}%
Equation for the correcting term $z_{1}$ is of form: 
\end{subequations}
\begin{equation}
\mathcal{\hat{L}}\left( \tfrac{d^{2}z_{1}}{dt^{2}}+\Omega ^{2}z_{1}\right) =%
\tfrac{3}{4}H\Omega A^{3}\sin \left( \Phi \right) +\tfrac{1}{4}A^{3}\left(
\left( 3+a\right) \Omega ^{2}-3b\right) \cos \left( \Phi \right) ,
\label{z1}
\end{equation}%
where $\Phi \equiv 3\Omega \tau +3\varphi \left( \Omega \right) $. Solving
Eqn.(\ref{z1}) and substituting to (\ref{sol1:1}) we get finally:%
\begin{equation}
z=A\left( \Omega \right) \cos \left( \Omega \tau +\varphi \right) -\tfrac{1}{%
32}A^{3}\left( \Omega \right) b\Omega \sin \left( \Phi \right) +\tfrac{1}{%
32\Omega ^{2}}A^{3}\left( \Omega \right) \cos \left( \Phi \right)   \label{z}
\end{equation}%
where $A\left( \Omega \right) ,\ \varphi \left( \Omega \right) $ are given
by Eqns.(\ref{AF}).

\section{General properties of the function $A\left( \Omega \right) $}

After introducing new variables, $\Omega ^{2}=X$, $A^{2}=Y$, the equation (%
\ref{AOmega}) defining the amplitude profile reads%
\begin{equation}
\begin{array}{l}
L\left( X,Y;a,b,h,H,\kappa ,J\right) \overset{df}{=}XY\left( pX-q-\tfrac{3}{4%
}Y\left( H-abX+bX^{2}\right) \right) ^{2} \\ 
+Y\left( X^{2}-rX+a+\tfrac{3}{4}Y\left( a-X+bHX^{2}\right) \right)
^{2}-JX^{2}=0%
\end{array}
\label{LXY}
\end{equation}%
where, as before, $p=h+H\left( \kappa +1\right) $, $q=ah+H$, $r=hH+a\left(
\kappa +1\right) +1$. A new parameter $J$ is a renormalized $G$, $J=\left( 
\tfrac{\kappa }{\kappa +1}\right) ^{2}G^{2}$. To obtain the corresponding
expression for the effective equation (\ref{eff}) one can put $G=\gamma 
\tfrac{\kappa +1}{\kappa }$ so that $J=\gamma ^{2}$ and then $\kappa =0$.

Singular points of $L\left( X,Y\right) $ are computed from equations \cite%
{Wall2004}: 
\begin{subequations}
\label{SING1}
\begin{eqnarray}
L &=&0,  \label{Sing1a} \\
\tfrac{\partial L}{\partial X} &=&0,  \label{Sing1b} \\
\tfrac{\partial L}{\partial Y} &=&0.  \label{Sing1c}
\end{eqnarray}%
We can eliminate $J$\ from Eqns. (\ref{Sing1a}), (\ref{Sing1b}) computing $L-%
\frac{1}{2}X\tfrac{\partial L}{\partial X}=\frac{1}{32}YK$ where 
\end{subequations}
\begin{equation}
\begin{array}{l}
K=-18b^{2}H^{2}X^{4}Y^{2}-24a^{2}\left( \kappa +1\right)
XY+24bhH^{2}X^{3}Y-48bHX^{4}Y \\ 
-18aXY^{2}-48aXY+32HhX^{3}+32a^{2}-27b^{2}X^{5}Y^{2}+16H^{2}X \\ 
-32aX-32a^{2}\left( \kappa +1\right) X+32a\left( \kappa +1\right)
X^{3}+9H^{2}XY^{2}+24H^{2}XY \\ 
+48a^{2}Y+18a^{2}Y^{2}+48hbX^{4}Y-9a^{2}b^{2}X^{3}Y^{2}+36ab^{2}X^{4}Y^{2}
\\ 
+16a^{2}h^{2}X-16\left( \kappa +1\right) ^{2}H^{2}X^{3}+48bH\left( \kappa
+1\right) X^{4}Y \\ 
-32hH\left( \kappa +1\right)
X^{3}-48abhX^{3}Y+32X^{3}-16h^{2}X^{3}-32X^{4}+24X^{3}Y%
\end{array}
\label{K1}
\end{equation}%
to obtain simplified equations:%
\begin{eqnarray}
K &=&0,  \label{Sing2a} \\
\tfrac{\partial L}{\partial Y} &=&0,  \label{Sing2b}
\end{eqnarray}%
from which $X$, $Y$ can be computed as functions of parameters $a$, $b$, $h$%
, $H$, $\kappa $ and, finally, $J$ can be computed from the last equation

\begin{equation}
\tfrac{\partial L}{\partial X}=0.  \label{Sing2c}
\end{equation}

Equations (\ref{Sing2a}), (\ref{Sing2b}), (\ref{Sing2c}) are still very
complicated making analytical investigation virtually impossible. We shall
thus solve these equations numerically.

\section{\label{comp}Computational results}

In the present Section singular points of amplitude profiles -- solutions of
Eqns. (\ref{Sing2a}), (\ref{Sing2b}), (\ref{Sing2c}) -- are studied. More
exactly, resonance curves with one singular point, two singular points on
one curve, and with degenerate singular point are presented and
metamorphoses of bifurcation diagrams are shown.

\subsection{\label{1s}Amplitude profiles with one singular point}

We have computed singular points for the following values of control
parameters: $\kappa =0.05$, $b=-0.001$, $H=0.4$, $a=5$, $h=0.5$ obtaining
four physical solutions (i.e. with $X>0$, $Y>0$, $J>0$):

\bigskip

\begin{tabular}{|l|l|l|l|}
\hline
\multicolumn{4}{|l|}{Table $1$.} \\ \hline
$X$ & $Y$ & $J$ & $n$ \\ \hline
$2.\,170\,051\,157$ & $1.\,357\,255\,661$ & $1.\,656\,917\,694\,08$ & $1$ \\ 
\hline
$4.\,835\,083\,103$ & $4.\,192\,055\,014$ & $1.\,036\,434\,991\,78$ & $2$ \\ 
\hline
$2.\,798\,801\,078$ & $1.\,237\,140\,868$ & $1.\,814\,387\,388\,23$ & $3$ \\ 
\hline
$4.\,153\,001\,386$ & $4.\,680\,111\,331$ & $0.\,963\,352\,653\,58$ & $4$ \\ 
\hline
\end{tabular}

\bigskip

\begin{figure}[th!]
\center \includegraphics[width=10cm, height=8cm]{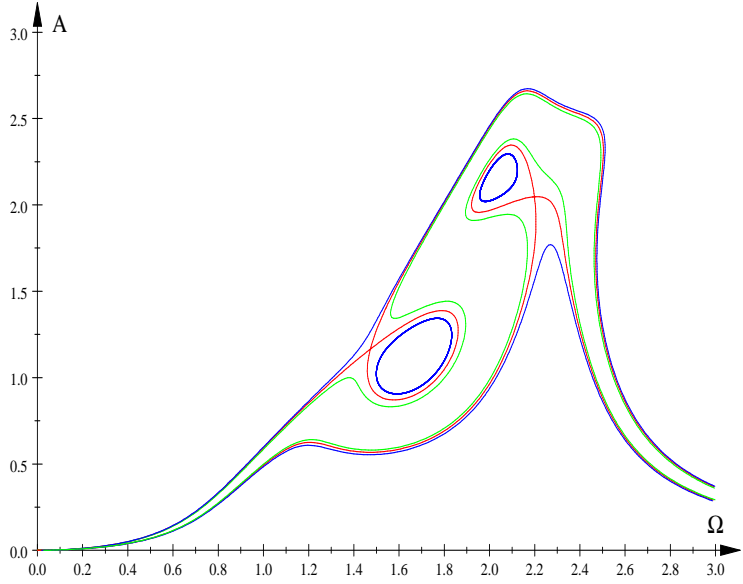}
\caption{Amplitude profiles with singular points, $\protect\kappa =0.05$, $%
b=-0.001$, $H=0.4$, $a=5$, $h=0.5$, $J=1.\,656\,917\,694\,08$ (left
self-intersection, red curve, $n=1$ in Table $1$), $J=1.\,036\,434\,991\,78$
(right self-intersection, red curve, $n=2$ in Table $1$) and neighbouring
curves (blue and green lines).}
\label{F1}
\end{figure}

The first two solutions correspond to self-intersections, see Fig. \ref{F1},
while the second pair represents isolated points.

Metamorphoses of bifurcation diagrams which occur in the neighbourhood of
self-intersections for the exact fourth-order equation are, for small $%
\kappa $, qualitatively similar to those studied for the case of $1:1$
resonance in the effective equation in \cite{Kyziol2011,Kyziol2012b} and are
not shown here.

\newpage

\subsection{\label{2s}Amplitude profiles with two singular points}

It is possible, tuning the parameters properly, to obtain amplitude profile
with two singular points.

Let, as before, $\kappa =0.05$, $b=-0.001$, $a=5$, $h=0.5$, $H$ being
arbitrary. We can compute, for some $H$, from (\ref{Sing2a}), (\ref{Sing2b}) 
$X\left( H\right) $, $Y\left( H\right) $, then from Eqn. (\ref{Sing2c}) we
get $J_{1}\left( H\right) $ and $J_{2}\left( H\right) $ corresponding to two
curves with one intersection each. The condition for a curve with two
intersections is $J_{1}=J_{2}$ for some $H$.

To find this value of $H$ we compute $J_{1}$, $J_{2}$ for two values of $H$, 
$H_{0}=0.40,$ \ $H_{1}=0.55,$ and use linear extrapolation to compute $%
H=H_{cr}$ such that $J_{1}\left( H_{cr}\right) =J_{2}\left( H_{cr}\right) $.
In one step of this procedure we compute new value of $H^{\left( i+2\right)
} $ from known $H^{\left( i\right) }$, $J_{1}^{\left( i\right) }$, $%
J_{2}^{\left( i\right) }$ and $H^{\left( i+1\right) }$, $J_{1}^{\left(
i+1\right) }$, $J_{2}^{\left( i+1\right) }$ solving linear system of
equations for $\alpha ^{\left( i,i+1\right) }$, $\beta ^{\left( i,i+1\right)
}$%
\begin{equation}
\begin{array}{rl}
J_{1}^{\left( i\right) }-J_{2}^{\left( i\right) } & =H^{\left( i\right)
}\alpha ^{\left( i,i+1\right) }+\beta ^{\left( i,i+1\right) } \\ 
J_{1}^{\left( i+1\right) }-J_{2}^{\left( i+1\right) } & =H^{\left(
i+1\right) }\alpha ^{\left( i,i+1\right) }+\beta ^{\left( i,i+1\right) }%
\end{array}
\label{extra}
\end{equation}%
where $i=0,1,2,\ldots $. Then the next value of $H^{\left( i+2\right) }$ is
computed as $H^{\left( i+2\right) }=-\dfrac{\beta ^{\left( i,i+1\right) }}{%
\alpha ^{\left( i,i+1\right) }}$. The convergence is quite fast, see Tables $%
2$, $3$.

\smallskip

\begin{tabular}{|l|l|l|l|l|}
\hline
\multicolumn{5}{|l|}{Table $2$} \\ \hline
$X_{1}^{\left( i\right) }$ & $Y_{1}^{\left( i\right) }$ & $J_{1}^{\left(
i\right) }$ & $H^{\left( i\right) }$ & $i$ \\ \hline
$2.\,170\,051\,157$ & $1.\,357\,255\,661$ & $1.\,656\,917\,694$ & $0.40$ & $%
0 $ \\ \hline
$2.\,222\,181\,140$ & $1.\,452\,883\,358$ & $1.\,736\,397\,285$ & $0.55$ & $%
1 $ \\ \hline
$2.\,250\,807\,092$ & $1.\,503\,767\,806$ & $1.\,775\,975\,009$ & $%
0.611\,498\,955$ & $2$ \\ \hline
$2.\,247\,348\,900$ & $1.\,497\,675\,627$ & $1.\,771\,343\,329$ & $%
0.604\,644\,433$ & $3$ \\ \hline
$2.\,247\,330\,153$ & $1.\,497\,642\,563$ & $1.\,771\,318\,111$ & $%
0.604\,606\,880\,$ & $4$ \\ \hline
$2.\,247\,330\,182$ & $1.\,497\,642\,613$ & $1.\,771\,318\,150$ & $%
0.604\,606\,937$ & $5$ \\ \hline
\end{tabular}

\medskip

\begin{tabular}{|l|l|l|l|l|}
\hline
\multicolumn{5}{|l|}{Table $3$} \\ \hline
$X_{2}^{\left( i\right) }$ & $Y_{2}^{\left( i\right) }$ & $J_{2}^{\left(
i\right) }$ & $H^{\left( i\right) }$ & $i$ \\ \hline
$4.\,835\,083\,103$ & $4.\,192\,055\,014$ & $1.\,036\,434\,992$ & $0.40$ & $%
0 $ \\ \hline
$4.\,793\,018\,228$ & $2.\,753\,798\,647$ & $1.\,555\,975\,411$ & $0.55$ & $%
1 $ \\ \hline
$4.\,620\,272\,267$ & $2.\,513\,582\,261$ & $1.\,798\,606\,879$ & $%
0.611\,498\,955$ & $2$ \\ \hline
$4.\,641\,565\,672$ & $2.\,538\,879\,342$ & $1.\,771\,466\,643$ & $%
0.604\,644\,433$ & $3$ \\ \hline
$4.\,641\,680\,711$ & $2.\,539\,018\,448$ & $1.\,771\,317\,923$ & $%
0.604\,606\,880$ & $4$ \\ \hline
$4.\,641\,680\,536$ & $2.\,539\,018\,236$ & $1.\,771\,318\,150$ & $%
0.604\,606\,937$ & $5$ \\ \hline
\end{tabular}

\smallskip

Figures below show convergence of two curves with one singular point to one
curve with two singular points.

\begin{figure}[th!]
\center \includegraphics[width=6cm, height=6cm]{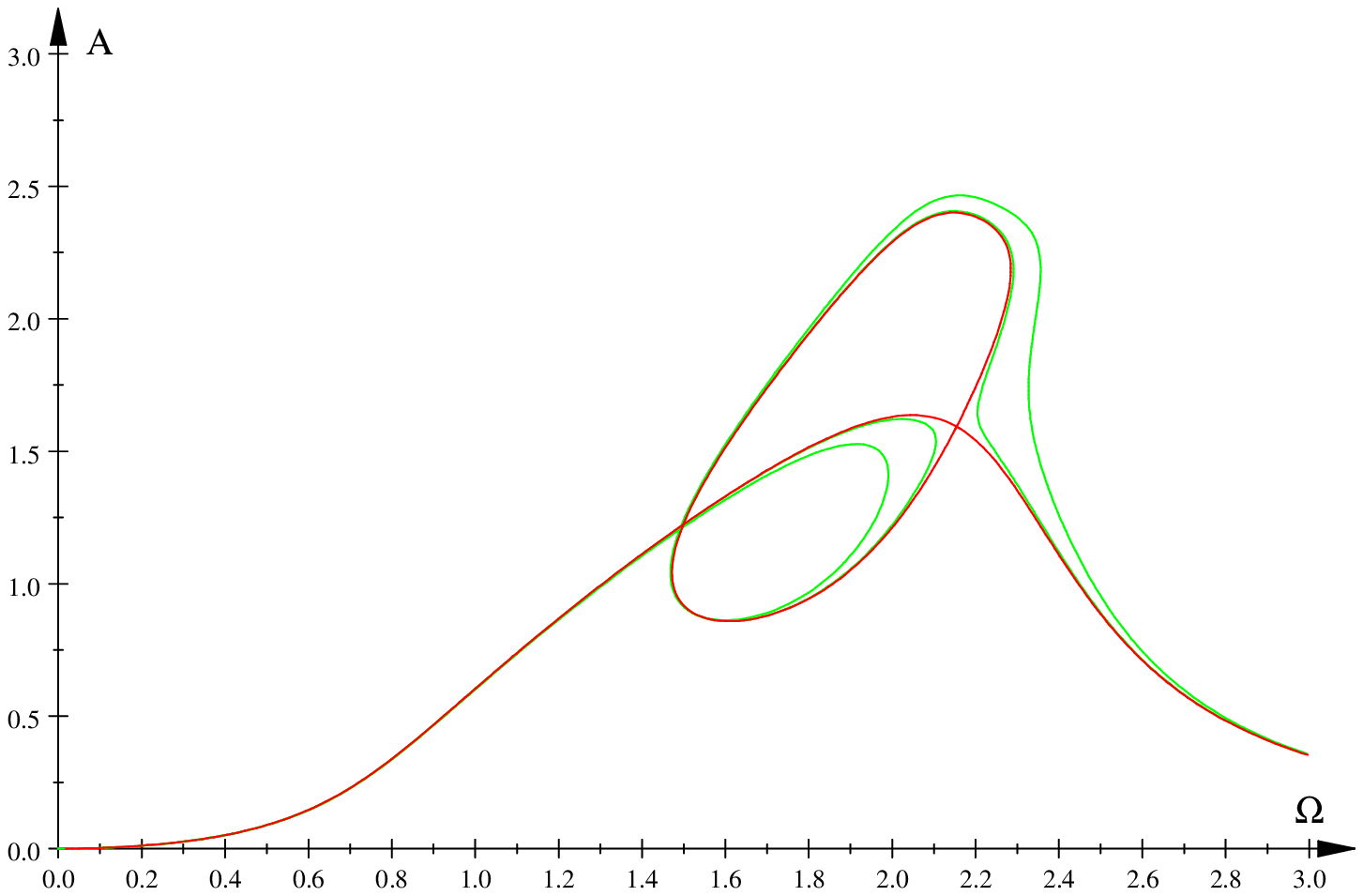} %
\includegraphics[width=6cm, height=6cm]{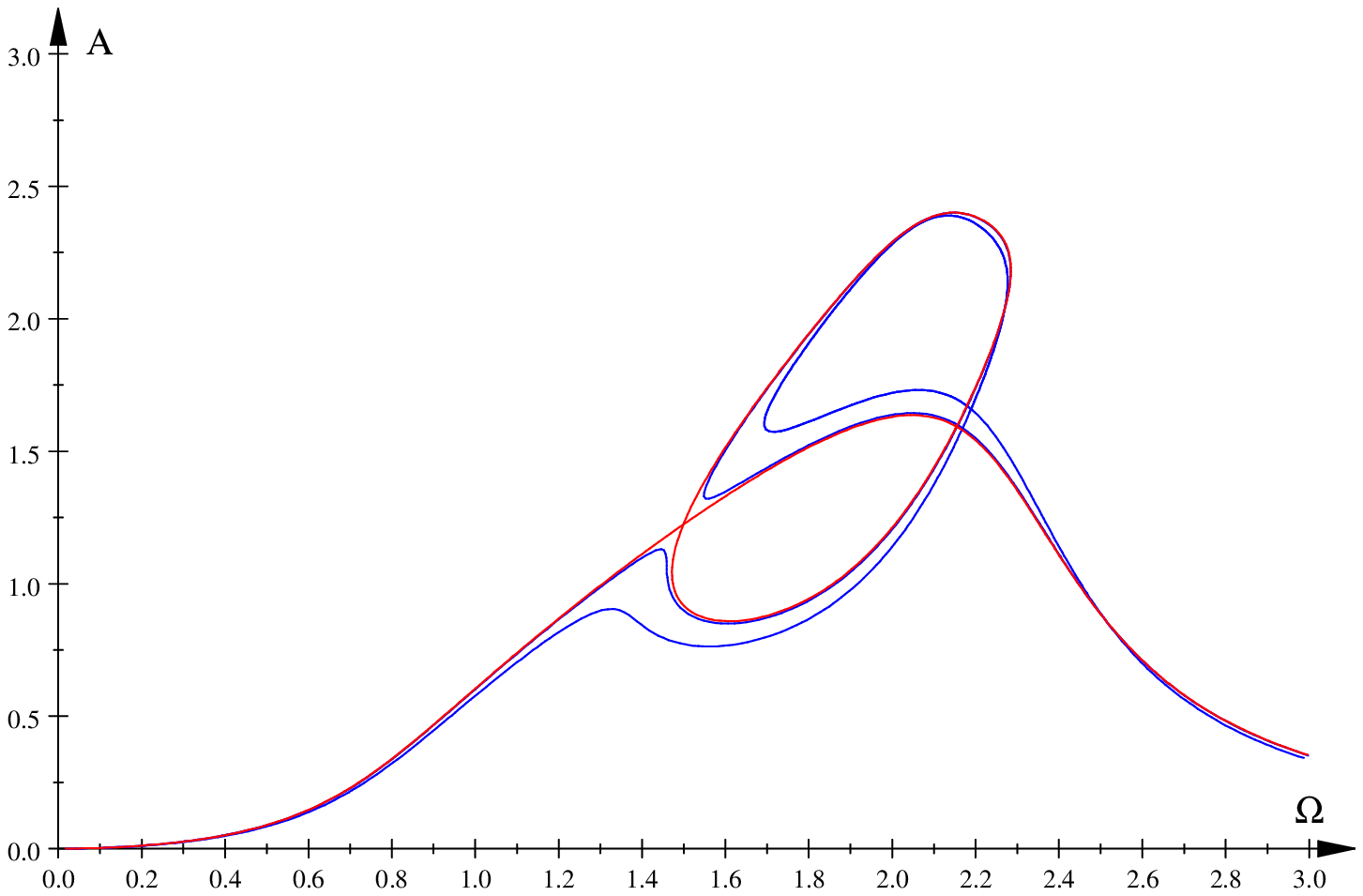}
\caption{Convergence of amplitude profiles to critical amplitude profile
with two singular points (red curve): convergence of curves from Table $2$
(left figure, green curves), and from Table $3$ (right figure, blue).}
\label{F2}
\end{figure}

\begin{figure}[th!]
\center \includegraphics[width=6cm, height=6cm]{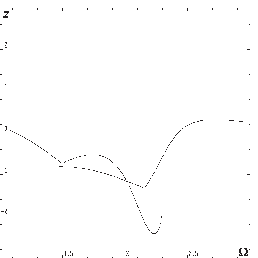} %
\includegraphics[width=6cm, height=6cm]{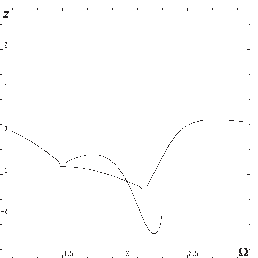}
\caption{Bifurcation diagrams. The left figure corresponds the amplitude
profile with singular point (left figure) with two cusps and to the
nonsingular curve (right figure) with two gaps.}
\label{F3}
\end{figure}

Bifurcation diagrams computed for parameters in the neighbourhood of such
resonance curve display presence of two singular points, see Figs. \ref{F3}.

In Figures \ref{F3} the parameters are $\kappa =0.05$, $b=-0.001$, $a=5$, $%
h=0.5$ in both cases and $J=1.\,745\,481\,261$, $H=0.6054$ for Fig. $3a$ and 
$J=1.\,740\,481\,261$, $H=0.6034$ for Fig. $3b$.

\newpage

\begin{figure}[th!]
\center \includegraphics[width=6cm, height=6cm]{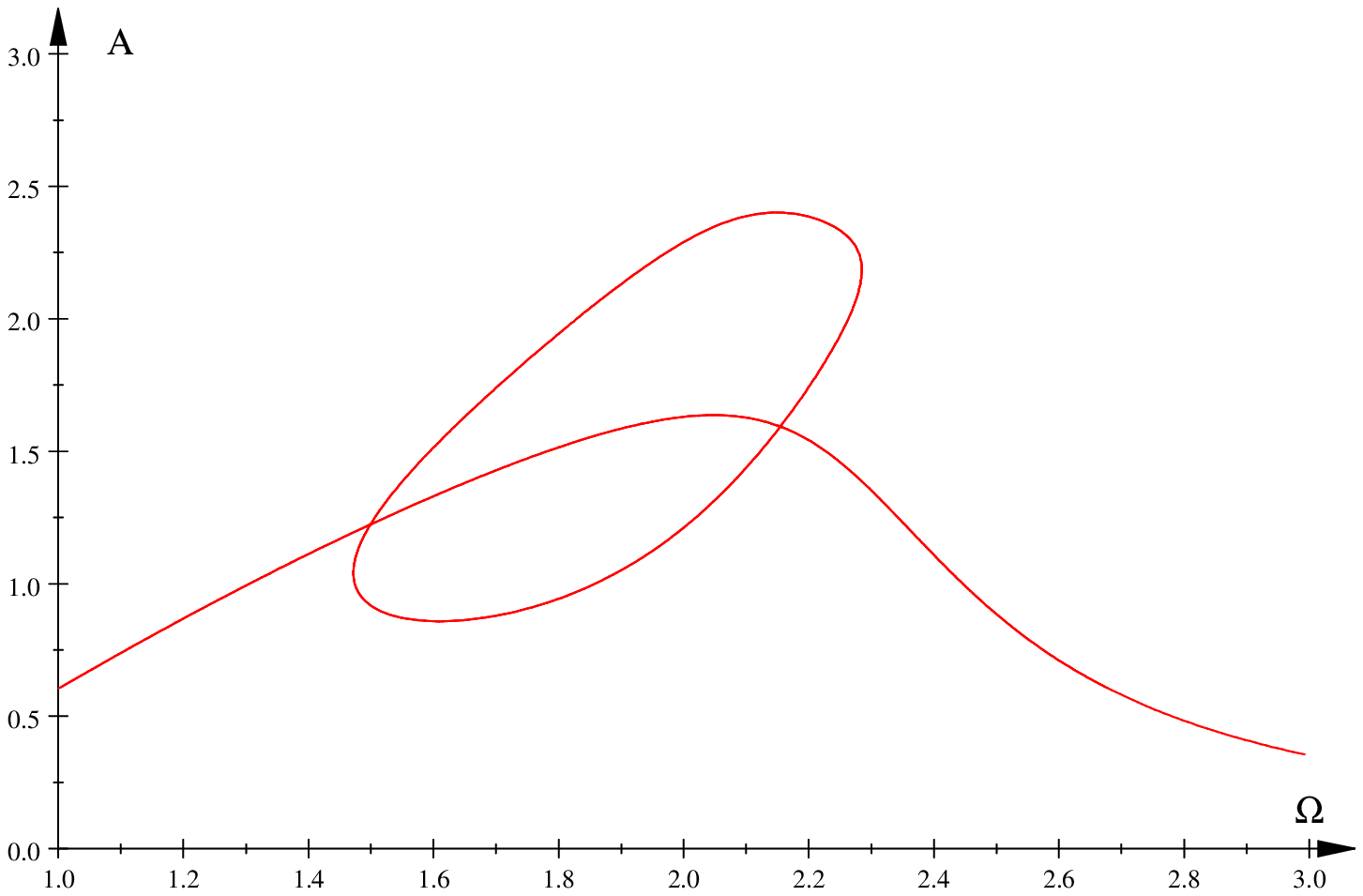} %
\includegraphics[width=6cm, height=6cm]{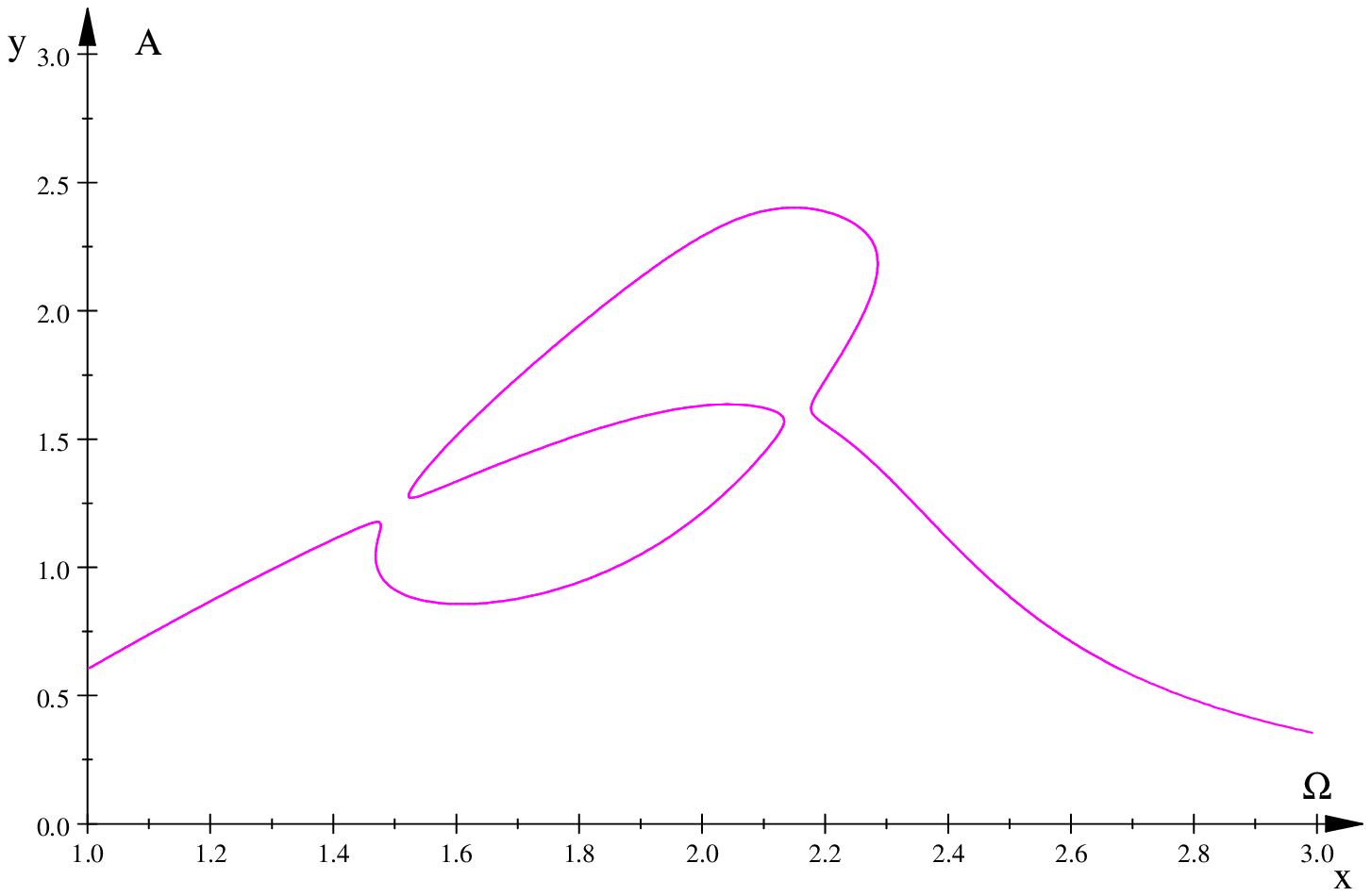}
\caption{The amplitude profile with two singular points (left figure, red)
and nonsingular curve (right figure, magenta).}
\label{F4}
\end{figure}

These diagrams correspond to amplitude profiles shown in Fig. \ref{F4}. 

\subsection{\label{1deg}Merging two singular points into a single degenerate
point}

It is possible, by smooth change of the parameters, to merge two singular
points lying on the red curve in Figs. $2$, $3$. The resulting singular
point is degenerate, i.e. fulfills the following set of equations \cite%
{Wall2004}:%
\begin{equation}
\begin{array}{lll}
L=0, & \frac{\partial L}{\partial X}=0, & \frac{\partial L}{\partial Y}%
=0,\medskip \\ 
\frac{\partial ^{2}L}{\partial X^{2}}=0, & \frac{\partial ^{2}L}{\partial
X\partial Y}=0, & \frac{\partial ^{2}L}{\partial Y^{2}}=0,%
\end{array}
\label{deg1}
\end{equation}%
where $L\left( X,Y\right) $ is given by (\ref{LXY}).

\begin{figure}[th!]
\center \includegraphics[width=8cm, height=6cm]{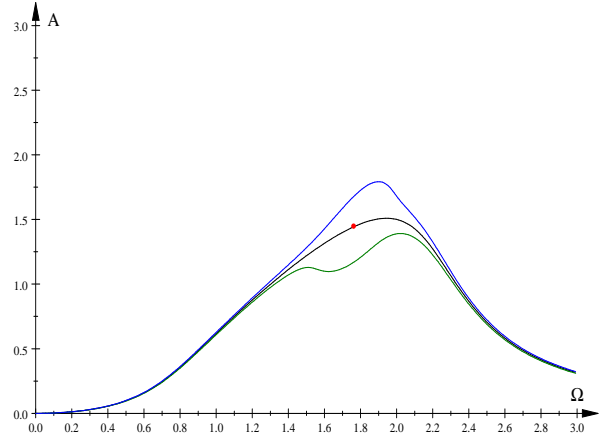}
\caption{Amplitude profile with degererate singular point (red dot) and two
neighbouring curves.}
\label{F5}
\end{figure}

\begin{figure}[th!]
\center \includegraphics[width=6cm, height=6cm]{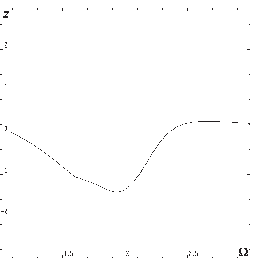} %
\includegraphics[width=6cm, height=6cm]{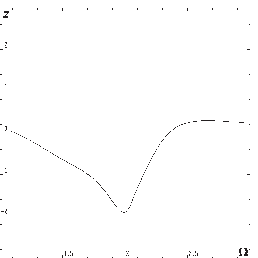}
\caption{Metamorphosis of the bifurcation diagrams near amplitude profile
with degenerate singular point.}
\label{F6}
\end{figure}

Solving Eqns. (\ref{deg1}) for $\kappa =0.05$, $J=1.\,771\,318\,150$ (these
two parameters correspond to the parameters of the critical red curve with
two singular points) we get $X=3.\,113\,090\,974$, $Y=2.\,087\,620\,813$, $%
h=0.\,548\,982\,679$, $a=4.\,538\,990\,962$, $b=-1.\,718\,542\,532\times
10^{-2}$, $H=0.\,644\,095\,068$, see Fig. \ref{F5}.

Bifurcation diagrams computed in the neighbourhood of the degenerate
singular point depend sensitively on small changes of parameters, Fig. \ref%
{F6}.

\section{Summary and discussion}

In this work we have studied dynamics of two coupled periodically driven
oscillators. The inner motion of this system has been described by the exact
fourth-order equation (\ref{4th-a}) (or (\ref{4th-b}) in nondimensional
form). Applying the KBM method we have computed approximate resonance curves
(amplitude profiles) $A\left( \Omega \right) $. Although the KBM method is
basically used for the second-order equations we managed to apply it to the
fourth-order equation since it was possible to eliminate secular terms and
impose steady-state conditions. Dependence of the amplitude $A$ on the
forcing frequency $\Omega $ is complex since $A\left( \Omega \right) $ is
defined implicitly as an algebraic curve, $L\left( X,Y\right) =0$, see Eqn.(%
\ref{LXY}), with polynomial function $L$ depending on variables $X=\Omega
^{2}$, $Y=A^{2}$ and control parameters $a$, $b$, $h$, $H$, $\kappa $, $J$
in a complicated manner.

In our previous paper we stressed that near singular points of algebraic
curves, defining amplitude profiles, metamorphoses of bifurcation diagrams
(and hence of dynamics) take place. In the present paper we have studied
three cases of singular points of the resonance curves defined by Eqn.(\ref%
{LXY}): i) the case of one singular point (Section \ref{1s}), ii) the case
of two singular points on one resonance curve (Section (\ref{2s})), iii) the
case of degenerate singular point (Section (\ref{1deg})). Indeed, dynamics
of the system (\ref{4th-b}) changes significantly in the neighbourhood of
singular points of resonance curve $L\left( X,Y\right) =0$. Singular points
described in Section \ref{comp} are just the tip of the iceberg and thus we
are going to study multitude of singular points of amplitude profiles (\ref%
{LXY}) in our future work.

\end{document}